\documentclass[aps,prd,preprint,groupedaddress]{revtex4}

\usepackage{bm}
\usepackage{slashed}
\usepackage{url}
\usepackage{float}

\usepackage{amsthm}

\newcommand{\beq}{\begin{eqnarray}}
\newcommand{\eeq}{\end{eqnarray}}

\newcommand{\bmp}{\noindent\begin{minipage}{16cm}}
\newcommand{\emp}{\end{minipage}\vskip 7mm} % 7mm untightened

\newcommand{\drawsquare}[2]{\hbox{%
\rule{#2pt}{#1pt}\hskip-#2pt% left vertical
\rule{#1pt}{#2pt}\hskip-#1pt% lower horizontal
\rule[#1pt]{#1pt}{#2pt}}\rule[#1pt]{#2pt}{#2pt}\hskip-#2pt%upper horizontal
\rule{#2pt}{#1pt}}% right vertical
\newcommand{\Yfund}{\raisebox{-.5pt}{\drawsquare{6.5}{0.4}}}% fund
\newcommand{\Ysymm}{\Yfund\hskip-0.4pt%
                    \Yfund}% symmetric second rank
\def\symm{\Ysymm}

\def\drawbox#1#2{\hrule height#2pt
        \hbox{\vrule width#2pt height#1pt \kern#1pt
              \vrule width#2pt}
              \hrule height#2pt}

\def\Asym#1#2{\vcenter{\vbox{\drawbox{#1}{#2}
              \kern-#2pt % line up boxes
              \drawbox{#1}{#2}}}}

\def\asymm{\Asym{6.4}{0.3}}

\begin{document}

% Use the \preprint command to place your local institutional report
% number in the upper righthand corner of the title page in preprint mode.
% Multiple \preprint commands are allowed.
% Use the 'preprintnumbers' class option to override journal defaults
% to display numbers if necessary
%\preprint{}

%Title of paper
\title{Parameters of NJL models for a generic representation of the gauge group}

% repeat the \author .. \affiliation  etc. as needed
% \email, \thanks, \homepage, \altaffiliation all apply to the current
% author. Explanatory text should go in the []'s, actual e-mail
% address or url should go in the {}'s for \email and \homepage.
% Please use the appropriate macro foreach each type of information

% \affiliation command applies to all authors since the last
% \affiliation command. The \affiliation command should follow the
% other information
% \affiliation can be followed by \email, \homepage, \thanks as well.
\author{Marco Frasca}\email{marcofrasca@mclink.it}
\affiliation{via Erasmo Gattamelata, 3, 00176 Roma, Italy}
%\homepage[]{Your web page}
%\thanks{}
%\altaffiliation{}

%Collaboration name if desired (requires use of superscriptaddress
%option in \documentclass). \noaffiliation is required (may also be
%used with the \author command).
%\collaboration can be followed by \email, \homepage, \thanks as well.
%\collaboration{}
%\noaffiliation

\date{\today}

\begin{abstract}
We generalize a non-local Nambu-Jona-Lasinio model to a generic representation of the gauge group. The critical temperature is given in a closed form as a function of the parameters of the theory and the cut-off. This result is generally useful in the understanding of QCD-like theories and their thermodynamical behavior.
\end{abstract}

% insert suggested PACS numbers in braces on next line
\pacs{}
% insert suggested keywords - APS authors don't need to do this
%\keywords{}

%\maketitle must follow title, authors, abstract, \pacs, and \keywords
\maketitle

% body of paper here - Use proper section commands
% References should be done using the \cite, \ref, and \label commands

QCD is widely accepted as the correct theory describing strong interacting matter. But low energy behavior entails great difficulties to be managed due to its non-perturbative nature. This implies a number of serious problems to understand a full thermodynamical behavior for the vacuum of the theory. So, a different way to overcome such difficulties has been devised in the use of QCD-like theories were an adjoint representation of Fermions is considered as a fundamental one \cite{arXiv:1005.2928,arXiv:1008.4145,arXiv:0911.2696}. This kind of theories find their most natural place in the technicolor approach to the Standard Model \cite{arXiv:1104.1255}.

A fundamental difficulty that arises in this case is that one just formulates a Nambu-Jona-Lasinio (NJL) model but has to postulate the corresponding parameters of the theory relying on plausibility arguments. Quite recently, it has been possible to give a clear proof of the fact that a NJL model is indeed the low-energy limit of QCD and all the parameters are consequently fixed through the parameters of QCD itself \cite{arXiv:1005.0314,arXiv:1105.5274,arXiv:0803.0319}, even if a better understanding of the approximations involved is in need. In \cite{arXiv:1105.5274} we were able to derive a non-local NJL model, founded on a preceding work \cite{arXiv:0810.1099}, proving that the instantons liquid model for the vacuum of QCD \cite{hep-ph/9610451} is an excellent approximation. Then, we obtained an expression for the critical temperature at zero quark mass and chemical potential that perfectly agrees with lattice computations provided the cut-off of the NJL model is taken to be about 770 MeV, a meaningful physical value for this model. The critical temperature obtained in this way agrees very well with a preceding theoretical derivation \cite{hep-ph/0410262} but we were able to give an explicit value to the mass gap of the theory obtaining a numerical evaluation.

The aim of this paper is to repeat all this for a generic QCD-like model and, in the end, to obtain the value of the cut-off for the critical temperature of about 2 GeV as generally found in literature \cite{hep-lat/9812023}.

Firstly, I point out some notational matter. We use the following table taken from \cite{arXiv:0711.3745} 
\begin{table}[H]
\begin{center}
    %\begin{minipage}{3.8in}
    \begin{tabular}{c||ccc }
    r & $ \quad T(r) $ & $\quad C_2(r) $ & $\quad
d(r) $  \\
    \hline \hline
    $ F $ & $\quad \frac{1}{2}$ & $\quad\frac{N^2-1}{2N}$ &\quad
     $N$  \\
        $\text{$A$}$ &\quad $N$ &\quad $N$ &\quad
$N^2-1$  \\
        $\symm$ & $\quad\frac{N+2}{2}$ &
$\quad\frac{(N-1)(N+2)}{N}$
    &\quad$\frac{N(N+1)}{2}$    \\
        $\asymm$ & $\quad\frac{N-2}{2}$ &
    $\quad\frac{(N+1)(N-2)}{N}$ & $\quad\frac{N(N-1)}{2}$
    \end{tabular}
    %\end{minipage}
    \end{center}
\caption{Relevant group factors for the representations as taken from
\cite{arXiv:0711.3745}.}\label{factors}
\end{table}
given the generators $T_r^a,\, a=1\ldots N^2-1$ of the gauge group in the
representation $r$, normalized as
$\text{Tr}\left[T_r^aT_r^b \right] = T(r) \delta^{ab}$. The
quadratic Casimir $C_2(r)$ is given by $T_r^aT_r^a = C_2(r)I$. Here
$C_2(r) d(r) = T(r) d(A)$ holds being $d(r)$ the
dimension of the representation $r$.

Then, the results given in \cite{arXiv:1105.5274} can be immediately generalized provided that:
\begin{enumerate}
\item {\sl 't Hooft coupling in a given representation is $d(r)g^2$.}
\item {\sl Number of components for the given representation $2NC_2(r)$.}
\end{enumerate}

With these statements the gluon propagator in the Landau gauge is given by\cite{arXiv:1105.5274}:
\begin{eqnarray}
\label{eq:Dmunu}
   D_{\mu\nu}^{ab}(p)&=&\delta_{ab}\left(\eta_{\mu\nu}-\frac{p_\mu p_\nu}{p^2}\right)\Delta(p^2)+O(1/\sqrt{d(r)}g)
\end{eqnarray}
being
\begin{equation}
\label{eq:gluonD}
    \Delta(p)=\sum_{n=0}^\infty\frac{B_n}{p^2-m_n^2+i\epsilon}
\end{equation}
and
\begin{equation}
    B_n=(2n+1)\frac{\pi^2}{K^2(i)}\frac{(-1)^{n+1}e^{-(n+\frac{1}{2})\pi}}{1+e^{-(2n+1)\pi}}.
\end{equation}
where $K(i)=\int_0^{\frac{\pi}{2}}\frac{d\theta}{\sqrt{1+\sin^2\theta}}\approx 1.3111028777$. The formula for the spectrum of the theory is
\begin{equation}
    m_n^{(r)} = \left(n+\frac{1}{2}\right)\frac{\pi}{K(i)}\left(\frac{d(r)g^2}{2}\right)^{\frac{1}{4}}\Lambda_r.
\end{equation}
From the mass spectrum we can identify a string tension that will be useful in the following. We can define
\begin{equation}
\label{eq:sigma}
    \sqrt{\sigma_r}=\left(\frac{d(r)g^2}{2}\right)^{\frac{1}{4}}\Lambda_r.
\end{equation}
Here $\Lambda_r$ is an arbitrary parameter arising from the integration of the equations of the theory and depending on the chosen representation of SU(N).

These results show without any doubt that a change of representation in the gauge group produces a difference in the low-energy behavior of the theory. Then, we can use them to fix the Nambu-Jona-Lasinio constant by noting that the non local form factor is \cite{arXiv:1105.5274}
\begin{equation}
   {\cal G}_r(p)=-\frac{1}{2}g^2\sum_{n=0}^\infty\frac{B_n}{p^2-(2n+1)^2(\pi/2K(i))^2\sigma_r+i\epsilon}=\frac{G_r}{2}{\cal C}_r(p)
\end{equation}
being $G_r$ the Nambu-Jona-Lasinio constant in the given representation that in our case is given by $G_r=2{\cal G}_r(0)=(g^2/\sigma_r)\sum_{n=0}^\infty\frac{B_n}{(2n+1)^2(\pi/2K(i))^2}\approx 0.7854(g^2/\sigma_r)$, so that ${\cal C}_r(0)=1$, definitely fixed by QCD. The value of $\sigma_r$ is given in eq.(\ref{eq:sigma}). So, finally we can formulate the NJL model for the adjoint representation as \cite{arXiv:1105.5274}
\begin{eqnarray}
   S_{QCD}^{(r)}&=&\frac{1}{2}\int d^4x\left[\frac{1}{2}(\partial\sigma)-\frac{1}{2}m_0^2\sigma^2\right]
   +\int d^4x\sum_q\bar q(x)i\slashed{\partial}q(x) \nonumber \\  
   &&-g^2\Lambda_r\int d^4xd^4y'\Delta(x-y')\sum_q\bar q(x)T^a_r\gamma^\mu q(x)
   \sum_{q'}\bar q'(y')T^a_r\gamma_\mu q'(y') \nonumber \\
   &&+O(1/\sqrt{d(r)}g).
\end{eqnarray}
The constant $\Lambda_r$ can be fixed through the value of the critical temperature for each representation as we will do in a moment. From this it is clear that the following holds
\begin{equation}
   \frac{G_A}{G_F}=\frac{\sigma_F}{\sigma_A}=\sqrt{\frac{d(F)}{d(A)}}\frac{\Lambda_F}{\Lambda_A}=\sqrt{\frac{N}{N^2-1}}\frac{\Lambda_F}{\Lambda_A}.
\end{equation}
The condition $\Lambda_A=\Lambda_F$ should be eventually proved with lattice computations or just guessed. Then, we can immediately put down the equation for the critical temperature for this model coming from the gap equation as\cite{arXiv:1105.5274}
\begin{equation}
   \label{eq:v}
    \frac{4d(r)N_f}{m_0^{(r)2}+1/G_r}T_{c-r}\sum_{k=-\infty}^\infty\int\frac{d^3p}{(2\pi)^3}{\cal C}^2_r(\omega_k,{\bm p})\frac{1}
    {\omega_k^2+{\bm p}^2}=1.
\end{equation}
being $m_0^{(r)}\approx 1.19\sqrt{\sigma_r}$, $G_r\approx 0.7854(g^2/\sigma_r)$ as seen above and the Matsubara frequencies are $\omega_k=(2n+1)\pi T_{c-r}$. The cut-off $\Lambda_{NJL-r}$ of the integral is the physical parameter to be computed given $T_{c-r}$. In order to evaluate the integral, we just note that the above equation can be rewritten as (in the Euclidean metric)
\begin{eqnarray}
   \label{eq:v2}
    &&\frac{4d(r)N_f}{m_0^{(r)2}+1/G_r}T_{c-r}\frac{1}{2\pi^2}\frac{g^4}{G_r^2}\sum_{k=-\infty}^\infty
    \sum_{n_1=0}^\infty\sum_{n_2=0}^\infty B_{n_1}B_{n_2}\times \\ \nonumber
    &&\int_0^\Lambda dpp^2\frac{1}
    {\omega_k^2+{\bm p}^2}\frac{1}{\omega_k^2+{\bm p}^2+m_{n_1}^2}\frac{1}{\omega_k^2+{\bm p}^2+m_{n_2}^2}=1.
\end{eqnarray}
The integral can be evaluated to give
\begin{eqnarray}
   I(\omega_k,n_1,n_2)&=&\left((\omega_k^2+m_{n_2}^2)^\frac{1}{2}\arctan\left(\frac{\Lambda}{(\omega_k^2+m_{n_2}^2)^\frac{1}{2}}\right)
   m_{n_1}^2(\omega_k^2+m_{n_1}^2)^\frac{1}{2}\right. \\ \nonumber
   &+&\omega_k\arctan\left(\frac{\Lambda}{\omega_k}\right)(\omega_k^2+m_{n_1}^2)^\frac{1}{2}(m_{n_2}^2-m_{n_1}^2) \\ \nonumber
   &-&\arctan\left(\frac{\Lambda}{(\omega_k^2+m_{n_1}^2)^\frac{1}{2}}\right)\omega_k^2m_{n_2}^2 \\ \nonumber
   &-&\left.\arctan\left(\frac{\Lambda}{(\omega_k^2+m_{n_1}^2)^\frac{1}{2}}\right)m_{n_1}^2m_{n_2}^2\right)
   \frac{1}{m_{n_2}^2(m_{n_1}^2-m_{n_2}^2)m_{n_1}^2
   (\omega_k^2+m_{n_1}^2)^\frac{1}{2}}.
\end{eqnarray}
Now, we assume that the critical temperature is much higher than the mass terms and we are able to approximate the above integral as
\begin{equation}
   I(\omega_k,n_1,n_2)\approx I(\omega_k,n_1,n_1) \approx \frac{\Lambda^3}{3\omega_k^6}
\end{equation}
where use has been made of the identity $\sum_{n=0}^\infty B_n=1$. Then, we can give the critical temperature in a closed form as
\begin{equation}
    T_{c-r}^5\approx\frac{1}{720\pi^2}\frac{d(r)N_f}{m_0^{(r)2}+1/G_r}\frac{g^4}{G_r^2}\Lambda^3
\end{equation}
using the fact that $\sum_{n=-\infty}^\infty 1/(2n+1)^6=\pi^6/480$. Our aim is to obtain the cut-off $\Lambda$ once the critical temperature is given. We assume that, as far as basic parameters of the gluonic part are considered, we can take a conservative step assuming it to be unchanged through different representations. So, $\Lambda_r=\Lambda_F=\Lambda_A=\ldots$ and the same for the coupling that we take $g\approx 3$ and $\sigma_F\approx (0.44)^2\ GeV^2$  for SU(3). Then, $\sigma_A=\sqrt{\frac{N^2-1}{N}}\sigma_F$ and the mass gap moves to higher energies, $m_0^{(A)}\approx 0.67\ GeV$, while NJL coupling is lowering. With these values, for SU(3) with two flavors and taking $T_{c-A}=2\ GeV$ one has immediately $\Lambda_{NJL-A}\approx 34\ GeV$ for the adjoint representation. It is easy to see that $m_0^{(A)}/\Lambda_{NJL-A}\approx 0.02$.

For aims of completeness, we extend this numerical study to the sextet case (symmetric representation). We will get $\sigma_S=\sqrt{\frac{N+1}{2}}\sigma_F$. This means that $m_0^{(S)}\approx 0.62\ GeV$. For $g\approx 3$, fixing $T_{c-S}=0.7\ GeV$ one gets $\Lambda_{NJL-S}\approx 6.88\ GeV$. This gives $m_0^{(S)}/\Lambda_{NJL-S}\approx 0.09$.

We have shown how the parameters of a Nambu-Jona-Lasinio model in a generic representation of the gauge group can be obtained from QCD with the parameters properly fixed. This has permitted us to obtain the critical temperature for this case and so, to determine the physical cut-off of the theory. This should help to frame better further analysis in this line of research.

% If you have acknowledgments, this puts in the proper section head.
%\begin{acknowledgments}
% put your acknowledgments here.
I would like to thank Marco Ruggieri for suggesting this question to extend my previous analysis of the low-energy behavior of QCD at finite temperature.
%\end{acknowledgments}

% Create the reference section using BibTeX:
%\bibliography{basename of .bib file}

\end{document}